\documentclass[12pt]{iopart}
\expandafter\let\csname equation*\endcsname\relax 
\expandafter\let\csname endequation*\endcsname\relax 
\usepackage{graphicx,amssymb,amsmath,amsthm}
\begin{document}
\newtheorem{theorem}{Theorem}
\newtheorem{prop}{Proposition}
\newtheorem{cor}{Corollary}
\newtheorem{defn}{Definition}
\def\theequation{\arabic{section}.\arabic{equation}}
\title{On an elliptic extension of the Kadomtsev-Petviashvili equation}
\author{Paul Jennings$^1$ and Frank Nijhoff$^2$}
\address{Department of Applied Mathematics, University of Leeds, Leeds, United Kingdom. LS2 9JT}
\eads{\mailto{$^1$pauljennings2104@gmail.com}, \mailto{$^2$nijhoff@maths.leeds.ac.uk}}

\begin{abstract}
A generalisation of the Lattice Potential Kadomtsev-Petviashvili (LPKP) equation is presented, using the method of Direct Linearisation based on an elliptic Cauchy kernel. This yields a $3+1$-dimensional lattice system with one of the lattice shifts singled out. The integrability of the lattice system  is considered, presenting a Lax representation and soliton solutions. An associated continuous system is also derived, yielding a $3+1$-dimensional generalisation of the potential KP equation associated with an elliptic curve.
\end{abstract}
\maketitle

\section{Introduction}
\setcounter{equation}{0}
To our knowledge, there exist only four truly elliptic integrable lattice systems, in the sense that they are naturally associated with an elliptic curve. These are:
\begin{itemize}
\item The lattice Landau-Lifshitz equations, resulting from a discretisation of the Sklyanin Lax pair\cite{Landau-Lifschitz}.
\item Adler's lattice Krichever-Novikov system, resulting from the permutability condition of the B\"{a}cklund transformations of the Krichever-Novikov equation\cite{Adler}.
\item Adler and Yamilov's system, arising from the consideration of Darboux chains.\cite{Adler/Yamilov}
\item The elliptic KdV system, resulting from a discrete linearisation scheme with an elliptic Cauchy kernel.\cite{ellipticKdv}
\end{itemize}
Apart from these, there is also a $5$-point scalar equation arising from the permutability of the B\"{a}cklund transformations for the Landau-Lifshitz equation\cite{Adler2}, although it is not clear as to whether this constitutes a true discretisation of  the Landau-Lifshitz equation by itself.

We present an elliptic extension of the Lattice Potential Kadomtsev-Petviashvili (LPKP) equation\cite{LPKP2}. This is the first such elliptic lattice system to reside in higher dimensions. The system is naturally a $3+1$ dimensional system, as opposed to the $2+1$ that may be expected. We also present the system's continuous analogue, i.e. an elliptic extension of the continuous (potential) KP equation. The only example of such a KP type system in the continuous setting was given by Date, Jimbo and Miwa\cite{Date/Jimbo/Miwa}. There are, of course, various elliptic integrable systems of continuous type (including the original Krichever-Novikov equation, Landau-Lifshitz equation and various others) in $1+1$-dimensions, but we will not go into these in the present note.

The elliptic  lattice KP system, which we will refer to as Ell-dKP, is derived  through a direct linearisation scheme following a similar method used in \cite{ellipticKdv} , which is employing an infinite matrix structure based on an elliptic Cauchy kernel.This scheme proves to be a powerful tool with the ability to provide, amongst others, a Lax representation, B\"{a}cklund and Miura transformations and hierarchies of commuting flows. It relies upon two key objects:
\begin{itemize}
\item Linear dynamics (either discrete or continuous) residing in the plane wave factors, $\rho(k)$.
\item A Cauchy kernel, $\Omega(k,k')$, through which the connection with the non-linear objects arises.
\end{itemize}
The resulting integrable lattice system is a set of simultaneous equations involving several components, which can be presented in various ways. One way of expressing this integrable lattice system is as the following system of discrete equations:
\begin{subequations}\label{eq:Lattice KP}
\begin{align}
&\left(p-\widetilde{u}\right)\left(q-r+\widetilde{\dot{u}}-\widetilde{\widehat{u}}\right)+\left(q-\widehat{u}\right)\left(r-p+\widehat{\widetilde{u}}-\widehat{\dot{u}}\right)+\left(r-\dot{u}\right)\left(p-q+\dot{\widehat{u}}-\dot{\widetilde{u}}\right)\nonumber\\
&\hspace{60mm}=g\left(\widetilde{\widehat{s}}'\left(\widetilde{s}-\widehat{s}\right)+\widehat{\dot{s}}'\left(\widehat{s}-\dot{s}\right)+\dot{\widetilde{s}}'\left(\dot{s}-\widetilde{s}\right)\right),\label{eq:LPKP}\\
&\frac{\left(p+\dot{u}\right)\dot{\widetilde{s}}-\left(q+\dot{u}\right)\widehat{\dot{s}}+\widehat{\dot{w}}-\widetilde{\dot{w}}}{\dot{s}}+
\frac{\left(q+\widetilde{u}\right)\widehat{\widetilde{s}}-\left(r+\widetilde{u}\right)\widetilde{\dot{s}}+\widetilde{\dot{w}}-\widetilde{\widehat{w}}}{\widetilde{s}}\nonumber\\
&\hspace{60mm}+\frac{\left(r+\widehat{u}\right)\dot{\widehat{s}}-\left(p+\widehat{u}\right)\widehat{\widetilde{s}}+\widehat{\widetilde{w}}-\widehat{\dot{w}}}{\widehat{s}}=0,\label{eq:ellKP2}\\
&\frac{\left(p-\dot{\widetilde{u}}\right)\dot{s}'-\left(r-\dot{\widetilde{u}}\right)\widetilde{s}'+\dot{w}'-\widetilde{w}'}{\dot{\widetilde{s}}'}+
\frac{\left(r-\widehat{\dot{u}}\right)\widehat{s}'-\left(q-\widehat{\dot{u}}\right)\dot{s}'+\widehat{w}'-\dot{w}'}{\widehat{\dot{s}}'}\nonumber\\
&\hspace{60mm}+\frac{\left(q-\widetilde{\widehat{u}}\right)\widetilde{s}'-\left(p-\widehat{\widetilde{u}}\right)\widehat{s}'+\widetilde{w}'-\widehat{w}'}{\widehat{\widetilde{s}}'}=0,\\
&\left(p+u-\frac{\widetilde{w}}{\widetilde{s}}\right)\left(p-\widetilde{u}+\frac{w}{s}\right)\nonumber\\
&\hspace{10mm}=p^2+\left(\underline{\widetilde{U}}_{1,0}-\widetilde{U}_{0,1}\right)-\left(U_{1,0}-U_{0,1}\right)+\frac{\widetilde{w}}{\widetilde{s}}\left(\widetilde{u}-\underline{\widetilde{u}}\right)-\left(\frac{1}{\widetilde{s}\underline{s}'}+3e+g\widetilde{s}'s\right),\label{eq:curve1}\\
&\left(q+u-\frac{\widehat{w}}{\widehat{s}}\right)\left(q-\widehat{u}+\frac{w}{s}\right)\nonumber\\
&\hspace{10mm}=q^2+\left(\underline{\widehat{U}}_{1,0}-\widehat{U}_{0,1}\right)-\left(U_{1,0}-U_{0,1}\right)+\frac{\widehat{w}}{\widetilde{s}}\left(\widehat{u}-\underline{\widehat{u}}\right)-\left(\frac{1}{\widehat{s}\underline{s}'}+3e+g\widehat{s}'s\right),\label{eq:curve2}\\
&\left(r+u-\frac{\dot{w}}{\dot{s}}\right)\left(r-\dot{u}+\frac{w}{s}\right)\nonumber\\
&\hspace{10mm}=r^2+\left(\underline{\dot{U}}_{1,0}-\dot{U}_{0,1}\right)-\left(U_{1,0}-U_{0,1}\right)+\frac{\dot{w}}{\dot{s}}\left(\dot{u}-\underline{\dot{u}}\right)-\left(\frac{1}{\dot{s}\underline{s}'}+3e+g\dot{s}'s\right),\label{eq:curve3}\\
&s'\overline{w}=w'\overline{s},
\end{align}
\end{subequations}
with $e$ and $g$ are fixed parameters, effectively the moduli of an elliptic curve
\begin{equation}\label{eq:elliptic curve}
\Gamma:y^2=1/x+3e+gx.
\end{equation}
In terms of  notation to describe the lattice system, for simplicity we let $u=u(n,m,l,N)$ denote the dependent variable for lattice points labeled by the quadruplet $(n,m,l,N)\in \mathbb{Z}^4$. The variables $p$, $q$ and $r$ are the continuous lattice parameters, associated with the grid size in the lattice directions, given by the independent variables $n$, $m$ and $l$, respectively.  The fourth variable $N$ is singled out in that it has no associated lattice parameter, effectively resulting in a (\ref{eq:Lattice KP}) being a $3+1$-dimensional system. For shifts in these lattice directions, we use the notation
\begin{equation}
\widetilde{u}=u(n+1,m,l,N),\hspace{5mm}\widehat{u}=u(n,m+1,l,N),\hspace{5mm},\dot{u}=u(n,m,l+1,N),
\end{equation}
together with
\begin{equation}
\overline{u}=u(n,m,l,N+1),
\end{equation}
and therefore, the combined shifts mean the following
\begin{equation}
\begin{array}{ccc}
\widehat{\widetilde{u}}=u(n+1,m+1,l,N),&\dot{\widetilde{u}}=u(n+1,m,l+1,N),\\
\dot{\widehat{u}}=u(n,m+1,l+1,N),&\widehat{\overline{u}}=u(n,m+1,l,N+1),\\
\overline{\widetilde{u}}=u(n+1,m,l,N+1),&\dot{\overline{u}}=u(n,m,l+1,N+1),\\
\widehat{\dot{\widetilde{u}}}=u(n+1,m+1,l+1,N),&\widehat{\dot{\overline{u}}}=u(n,m+1,l+1,N+1),\\
\widehat{\overline{\widetilde{u}}}=u(n+1,m+1,l,N+1),&\overline{\dot{\widetilde{u}}}=u(n+1,m,l+1,N+1),\\
\overline{\widehat{\dot{\widetilde{u}}}}=u(n+1,m+1,l+1,N+1),& 
\end{array}
\end{equation}
completing the notation for the vertices of an elementary lattice octachoron.

Integrability of the system is understood in terms of the existence of a Lax representation and soliton solutions, both of which we derive later.

Equations (\ref{eq:Lattice KP}) is a system of seven interconnected equations for dependent variables $u$, $s$, $s'$, $w$, $w'$, $U_{0,1}$ and $U_{1,0}$, but the latter two are understood to be eliminated by pairwise combinations of (\ref{eq:curve1}, \ref{eq:curve2}, \ref{eq:curve3}). (We prefer to leave $U_{0,1}$ and $U_{1,0}$ in the system to make manifest the dependence on the modulus e, although we do include this other presentation later (\ref{eq:no e})). This system is a natural higher-dimensional extension of the lattice potential KP equation \cite{LPKP}, which can be recovered by setting $g=0$ in equation (\ref{eq:LPKP}), in which case the elliptic curve degenerates into a rational curve. On the other hand, the elliptic KdV equation \cite{ellipticKdv} may also be recovered through a dimensional reduction.

An analogous continuous system, associated with the same elliptic curve (\ref{eq:elliptic curve}), is given by
\begin{subequations}\label{eq:continuousKP}
\begin{align}
&\left(u_t-\frac{1}{4}u_{xxx}-\frac{3}{2}\left(u_x\right)^2+\frac{3}{2}gs_xs'_x\right)_x=\frac{3}{4}u_{yy}+\frac{3}{2}g\left(s'_xs_y-s'_ys_x\right),\label{eq:KP1}\\
&\left(ss'\right)_t=\frac{1}{4}\left(s_{xxx}s'+s'_{xxx}s\right)+\frac{3}{2}u_x\left(ss'\right)_x-3us_xs'_x+\frac{3}{4}\left(s's_{y}-ss'_{y}-s_xs'_x\right)_x\nonumber\\
&\hspace{40mm}+\frac{3}{2}\left(us's_y-uss'_y+w'_ys-w_ys'\right)+\frac{3}{2}\left(w_xs'_x+w'_xs_x\right),\\
&\left(ss'\right)_y=2sw'_x-2s'w_x+2u\left(s_xs'-s'_xs\right)+s_{xx}s'-s'_{xx}s,\label{eq:sy}\\
&\left(u+\frac{w}{s}\right)_x+\left(u-\frac{w}{s}\right)^2+\frac{w}{s}\left(u-\underline{u}\right)=\left(\frac{1}{\underline{s}'s}+3e+gss'\right)+U_{1,0}-\underline{U}_{1,0},\label{eq:curve dep1}\\
&\left(u+\frac{w'}{s'}\right)_x+\left(u-\frac{w'}{s'}\right)^2+\frac{w'}{s'}\left(u-\overline{u}\right)=\left(\frac{1}{\overline{s}'s}+3e+gss'\right)+U_{0,1}-\overline{U}_{0,1},\label{eq:curve dep2}\\
&s'\overline{w}=w'\overline{s},\label{eq:bar}
\end{align}
\end{subequations}
which, in fact, can be obtained from a continuum limit of the lattice system. Note that this system is dependent on $3$ continuous variables and one discrete variable. Here again, the elimination of the variables $U_{1,0}$ and $U_{0,1}$ disguises the dependence on the elliptic modulus $e$. This may be achieved by differentiating both equations (\ref{eq:curve dep1}) and (\ref{eq:curve dep2}) with respect to $x$, and substituting in equations (\ref{eq u10 diff}) and (\ref{eq u01 diff}), respectively. Again the degeneration of the elliptic curve results in potential KP equation(\ref{eq:KP1}), whilst the elliptic potential KdV equation [17]is recovered through a dimensional reduction.

In this paper, we will provide a derivation of both the continuous as well as the lattice elliptic KP systems, and establish some of their properties.

\section{Elliptic Matrix Structure}
\setcounter{equation}{0}
The starting point for the elliptic analogue to the lattice potential KP equation is the algebra of centred infinite elliptic matrices, $\mathcal{A}$. This is an associative algebra, with identity $\boldsymbol{1}$, which is quasigraded, with grading given by two types of raising operator, $\boldsymbol{\Lambda}$ of order one, and $\boldsymbol{L}$ of order two, such that
\begin{equation}\label{eq:curve}
\begin{array}{cc}
\boldsymbol{\Lambda}^2=\boldsymbol{L}+3e\boldsymbol{1}+g\boldsymbol{L}^{-1}, & \boldsymbol{L}\boldsymbol{\Lambda}=\boldsymbol{\Lambda}\boldsymbol{L},
\end{array}
\end{equation}
where $e,g\in \mathbb{C}$ are the moduli of the elliptic curve given by this relation\cite{ellipticKdv}.  We point out that, unlike ordinary matrices, these elliptic matrices cannot be written as simple arrays of entries in the conventional way, but are built from the action of two index raising operators $\boldsymbol{\Lambda}$ and $\boldsymbol{L}$ and their conjugates and projector $\boldsymbol{O}$. The projector $\boldsymbol{O}$ is the projection matrix on the central entry, with $\left(\boldsymbol{O}\cdot\boldsymbol{A}\right)_{i,j}=\delta_{i,0}\boldsymbol{A}_{0,j}$ and $\left(\boldsymbol{A}\cdot\boldsymbol{O}\right)_{i,j}=\delta_{0,j}\boldsymbol{A}_{i,0}$ for any $\boldsymbol{A}\in\mathcal{A}$. We adopt the following convention of labelling the entries of these elliptic matrices, $\boldsymbol{A}\in\mathcal{A}$, with $\boldsymbol{A}=\left(A_{i,j}\right)$ for $i,j\in\mathbb{Z}$, and central entry $A_{0,0}$ as follows:
\begin{subequations}
\begin{align}
A_{2i,2j}&=\left(\boldsymbol{L}^i\boldsymbol{A}{}^t\boldsymbol{L}^j\right)_{0,0},\\
A_{2i+1,2j}&=\left(\boldsymbol{L}^i\boldsymbol{\Lambda}\boldsymbol{A}{}^t\boldsymbol{L}^j\right)_{0,0},\\
A_{2i,2j+1}&=\left(\boldsymbol{L}^i\boldsymbol{A}{}^t\boldsymbol{\Lambda}{}^t\boldsymbol{L}^j\right)_{0,0},\\
A_{2i+1,2j+1}&=\left(\boldsymbol{L}^i\boldsymbol{\Lambda}\boldsymbol{A}{}^t\boldsymbol{\Lambda}{}^t\boldsymbol{L}^j\right)_{0,0}.
\end{align}
\end{subequations}
To make these elliptic matrices more explicit, we consider their action on infinite component vectors of the form
\begin{equation}
\boldsymbol{c}_{\kappa}=\left(\kappa^{\nu}\right)_{\nu\in\mathbb{Z}},
\end{equation}
i.e. infinite column vectors in a basis of monomials in a variables $\kappa$, which can be viewed as uniformising vectors for the curve $\Gamma$, (\ref{eq:elliptic curve}). The corresponding row vector is given by its transpose, ${}^t\boldsymbol{c}_{\kappa'}$. The crucial property of these vectors is that they are eigenvectors under the actions of $\boldsymbol{\Lambda}$, and  ${}^t\boldsymbol{\Lambda}$, 
\begin{equation}\label{eq:Lambda action}
\begin{array}{cc}
\boldsymbol{\Lambda}\boldsymbol{c}_{\kappa}=\kappa\boldsymbol{c}_{\kappa} & {}^t\boldsymbol{c}_{\kappa'}{}^t\boldsymbol{\Lambda}=\kappa'\boldsymbol{c}_{\kappa'},
\end{array}
\end{equation}
and the actions $\boldsymbol{L}$, and ${}^t\boldsymbol{L}$,
\begin{equation}\label{eq:L action}
\begin{array}{cc}
\boldsymbol{L}\boldsymbol{c}_{\kappa}=K\boldsymbol{c}_{\kappa}, & {}^t\boldsymbol{c}_{\kappa'}{}^t\boldsymbol{L}=L\boldsymbol{c}_{\kappa'},
\end{array}
\end{equation}
respectively. The eigenvalues form pairs $(k,K)$ and $(k',K')$, which are points on the elliptic curve $\Gamma$, (\ref{eq:elliptic curve})
\begin{equation}
k^2=K+3e+\frac{g}{K},\hspace{20mm}k'^2=K'+3e+\frac{g}{K'},
\end{equation}
whereby
\begin{subequations}
\begin{align}
k&=\frac{1}{2}\frac{\wp'\left(\kappa\right)}{\wp\left(\kappa\right)-e},\hspace{5mm} K=\wp\left(\kappa\right)-e,\\
k'&=\frac{1}{2}\frac{\wp'\left(\kappa'\right)}{\wp\left(\kappa'\right)-e},\hspace{5mm} K'=\wp\left(\kappa'\right)-e,\\
e&=\wp(\kappa).
\end{align}
\end{subequations}
For the sake of the construction of the integrable system, we introduce
\begin{equation}\label{eq:Cauchy1}
\boldsymbol{\Omega} \leftrightarrow \Omega\left(k,k'\right)=\frac{k-k'}{K-K'}=\frac{1-g/\left(KK'\right)}{k+k'},
\end{equation}
The Cauchy Kernel for the system, $\boldsymbol{\Omega}\in\mathcal{A}$,  then obeys
\begin{subequations}
\begin{align}
&\boldsymbol{\Omega}\boldsymbol{\Lambda}+{}^t\boldsymbol{\Lambda}\boldsymbol{\Omega}=\boldsymbol{O}-g{}^t\boldsymbol{L}^{-1}\boldsymbol{O}\boldsymbol{L}^{-1}=:\widehat{\boldsymbol{O}},\label{eq:Cauchy}\\
&\boldsymbol{\Omega}\boldsymbol{L}-{}^t\boldsymbol{L}\boldsymbol{\Omega}=\boldsymbol{O}\boldsymbol{\Lambda}-{}^t\boldsymbol{\Lambda}\boldsymbol{O},\\
&\boldsymbol{\Omega}\boldsymbol{\Lambda}^j-\left(-{}^t\boldsymbol{\Lambda}\right)^j\boldsymbol{\Omega}=\sum^{j-1}_{i=0}\left(-{}^t\Lambda\right)^i\boldsymbol{\widehat{O}}\boldsymbol{\Lambda}^{j-1-i}=:\boldsymbol{\widehat{O}}_j,\label{eq:Ojhat}\\
&\boldsymbol{\widehat{O}}_j\boldsymbol{\Lambda}^i+\left(-{}^t\boldsymbol{\Lambda}\right)^j\boldsymbol{\widehat{O}}_i=\boldsymbol{\widehat{O}}_{i+j},
\end{align}
\end{subequations}
The dynamics of the system are encoded by  $\boldsymbol{C}\in\mathcal{A}$, given by the formal integral
\begin{equation}\label{eq:C}
\boldsymbol{C}=\iint \limits_D \, \mathrm{d}\mu\left(\lambda,\lambda'\right)\,\rho_{\lambda}\textbf{c}_{\lambda}{}^t\textbf{c}_{\lambda'}\sigma_{\lambda'},
\end{equation}
over an arbitrary surface, $D$, on the space of variables $\lambda$, $\lambda'$. $\rho_{\kappa}$ and $\sigma_{\kappa}'$, the plane wave factors, are discrete exponential functions, given by 
\begin{subequations}
\begin{align}
&\rho_{\kappa}(n,N)=(p+k)^n(-K)^N\rho_{\kappa}(0,0),\\
&\sigma_{\kappa'}(n,N)=(p-k')^{-n}(-K')^{-N}\sigma_{\kappa'}(0,0),
\end{align}
\end{subequations}
respectively. The factors $\textbf{c}_k$ and ${}^t\textbf{c}_{k'}$ are the infinite component vectors defined earlier. The integration measure, $\mathrm{d} \mu\left(\lambda,\lambda'\right)$, is in principle arbitrary, but we assume that basic operations, such as differentiation, and shifts, with respect to the parameters, commute with the integrations. Lattice shifts of these plane wave factors result in
\begin{subequations}\label{eq:shift}
\begin{align}
&\widetilde{\boldsymbol{C}}\left(p-{}^t\boldsymbol{\Lambda}\right)=\left(p+\boldsymbol{\Lambda}\right)\boldsymbol{C}, \label{eq:tilde}\\
&\widehat{\boldsymbol{C}}\left(q-{}^t\boldsymbol{\Lambda}\right)=\left(q+\boldsymbol{\Lambda}\right)\boldsymbol{C}, \label{eq:hat''}\\
&\dot{\boldsymbol{C}}\left(r-{}^t\boldsymbol{\Lambda}\right)=\left(r+\boldsymbol{\Lambda}\right)\boldsymbol{C}, \label{eq:dot}\\
&-\overline{\boldsymbol{C}}{}^t\boldsymbol{L}=-\boldsymbol{L}\boldsymbol{C}. \label{eq:hat}
\end{align}
\end{subequations}
The main object from which the non-linear equations are obtained is the infinite matrix, $\boldsymbol{U}\in\mathcal{A}$, defined
\begin{equation}
U\equiv\boldsymbol{C}\left(\boldsymbol{1}+\boldsymbol{\Omega}\boldsymbol{C}\right)^{-1}\equiv\left(\boldsymbol{1}-\boldsymbol{U}\boldsymbol{\Omega}\right)\boldsymbol{C},
\end{equation} 
with components $U_{i,j}$, together with infinite component vectors, $\boldsymbol{u}_k$, defined through the equations,
\begin{subequations}
\begin{align}
&\boldsymbol{u}_k+\rho_k\boldsymbol{U}\boldsymbol{\Omega}\boldsymbol{c}_k=\rho_k\boldsymbol{c}_k,\label{eq:u vector}\\
&{}^t\boldsymbol{u}_{k'}+{}^t\boldsymbol{c}_{k'}\boldsymbol{\Omega}{}^t\boldsymbol{U}\sigma_{k'}={}^t\boldsymbol{c}_k\sigma_{k'}.
\end{align}
\end{subequations}
For this infinite matrix $\boldsymbol{U}$, discrete Riccati type shift relations follow from those for $\boldsymbol{C}$, (\ref{eq:shift}), giving
\begin{subequations}\label{U shifts}
\begin{align}
\widetilde{\boldsymbol{U}}\left(p-{}^t\boldsymbol{\Lambda}\right)&=\left(p+\boldsymbol{\Lambda}\right)\boldsymbol{U}-\widetilde{\boldsymbol{U}}\left(\boldsymbol{O}-g{}^t\boldsymbol{L}^{-1}\boldsymbol{O}\boldsymbol{L}^{-1}\right)\boldsymbol{U},\label{eq:tilde2}\\
\widehat{\boldsymbol{U}}\left(q-{}^t\boldsymbol{\Lambda}\right)&=\left(q+\boldsymbol{\Lambda}\right)\boldsymbol{U}-\widehat{\boldsymbol{U}}\left(\boldsymbol{O}-g{}^t\boldsymbol{L}^{-1}\boldsymbol{O}\boldsymbol{L}^{-1}\right)\boldsymbol{U},\label{eq:hat'2}\\
\dot{\boldsymbol{U}}\left(r-{}^t\boldsymbol{\Lambda}\right)&=\left(r+\boldsymbol{\Lambda}\right)\boldsymbol{U}-\dot{\boldsymbol{U}}\left(\boldsymbol{O}-g{}^t\boldsymbol{L}^{-1}\boldsymbol{O}\boldsymbol{L}^{-1}\right)\boldsymbol{U},\label{eq:dot2}\\
-\overline{\boldsymbol{U}}{}^t\boldsymbol{L}&=\-\boldsymbol{L}\boldsymbol{U}+\overline{\boldsymbol{U}}\left(\boldsymbol{O}\boldsymbol{\Lambda}-{}^t\boldsymbol{\Lambda}\boldsymbol{O}\right)\boldsymbol{U},\label{eq:barshift2}
\end{align}
\end{subequations}
respectively. We also have
\begin{equation}\label{eq:barshift1}
\boldsymbol{U}{}^t\boldsymbol{L}^{-1}=\boldsymbol{L}^{-1}\overline{\boldsymbol{U}}+\boldsymbol{U}{}^t\boldsymbol{L}^{-1}\left(\boldsymbol{O}\boldsymbol{\Lambda}-{}^t\boldsymbol{\Lambda}\boldsymbol{O}\right)\boldsymbol{L}^{-1}\overline{\boldsymbol{U}},
\end{equation}
which holds if and only if (\ref{eq:barshift2}) also does. Using these shift relations we are now able to derive a system of relations in terms of $\boldsymbol{U}$'s matrix entries.

\section{Elliptic Lattice Structure}
\setcounter{equation}{0}
Having obtained the basic relations in the previous section (\ref{eq:barshift1},\ref{U shifts}) in terms of the elliptic matrix $\boldsymbol{U}$, closed-form equations can now be derived in terms of a well-chosen set of entries. To do this we single out the following entries:
\begin{equation}
\begin{array}{ccc}
u=U_{0,0}, & s=U_{-2,0}, & s'=U_{0,-2},\\
h=U_{-2,-2}, & v=1-U_{-1,0}, & v'=1-U_{0,-1},\\
{} & w=1+U_{-2,1}, & w'=1+U_{1,-2},\\
\end{array}
\end{equation}
Starting with (\ref{eq:barshift1}), which involve only the $\overline{\cdot}$ shifts, we can derive the following set of relations:
\begin{subequations}
\begin{align}
&U_{1,-1} = \frac{1-\overline{v}w'}{\overline{s}},\label{eq:u1-1}\\
&\overline{U}_{-1,1}=\frac{1-\overline{v}w'}{s'},\label{eq:u-1,1}\\
&\overline{U}_{-4,0}=\overline{s}U_{-2,-1}+\overline{v}h,\label{eq:U_-4,0}\\
&U_{0,-4}=s'\overline{U}_{-1,-2}+v'\overline{h},\\
&\overline{U}_{-2,2}=u\overline{w}-\overline{s}U_{1,0},\\
&U_{2,-2}=\overline{u}w'-s'\overline{U}_{0,1},\\
&\frac{\overline{s}}{s'}=\frac{\overline{v}}{v'}=\frac{\overline{w}}{w'}=\frac{\overline{U}_{-1,1}}{U_{1,-1}}.
\end{align}
\end{subequations}
For the other lattice shift directions, we take as a starting point equation (\ref{eq:tilde2}), from which the following equations are derived:
\begin{subequations}
\begin{align}
&p\left(\widetilde{u}-u\right)+\widetilde{u}u=\widetilde{U}_{0,1}+U_{1,0}+g\widetilde{s}'s,\label{eq:u tilde}\\
&p-gh=\frac{\left(p-\widetilde{u}\right)s'+w'-\widetilde{v}'}{\widetilde{s}'},\label{eq:s' tilde}\\
&p+g\widetilde{h}=\frac{\left(p+u\right)\widetilde{s}+v-\widetilde{w}}{s},\label{eq:s tilde}\\
&U_{-1,-2}+\widetilde{U}_{-2,-1}=p\left(\widetilde{h}-h\right)-g\widetilde{h}h+\widetilde{s}s',\\
&p\left(v-\widetilde{v}\right)=u\widetilde{v}+\widetilde{U}_{-1,1}+gs\left(\widetilde{U}_{-1,-2}+\underline{U}_{-2,-1}\right)+3es+gv\underline{h},\\
&p\left(v'-\widetilde{v}'\right)=\widetilde{u}v'+U_{1,-1}+g\widetilde{s}'\left(U_{-2,-1}+\widetilde{\overline{U}}_{-1,-2}\right)+3e\widetilde{s}'+g\widetilde{v}'\widetilde{\overline{h}},\\
&p\left(\widetilde{w}-w\right)=\widetilde{\underline{u}}\widetilde{w}-\widetilde{s}\left(\widetilde{\underline{U}}_{1,0}+U_{0,1}\right)+U_{-1,1}+3e\widetilde{s}+g\widetilde{h}w,\label{eq:w tilde}\\
&p\left(\widetilde{w}'-w'\right)=\overline{u}w'-{s}'\left(\widetilde{U}_{1,0}+\overline{U}_{0,1}\right)+\widetilde{U}_{1,-1}+3es'+gh\widetilde{w}'.
\end{align}
\end{subequations}
These involve only the $\widetilde{\cdot}$ shifts, but similar relations are obtained in an obvious way by replacing $p$ wiith  $q$ and $r$, and the $\widetilde{\cdot}$ shifts with $\widehat{\cdot}$ and $\dot{\cdot}$ shifts, respectively. By combining the various relations, and eliminating the variables $h$, $v$, $v'$, $U_{-1,-2}$ and $U_{-2,-1}$, the closed-form system of partial difference equations given earlier (\ref{eq:Lattice KP}) can be found. For example, by eliminating $v$ and $h$ from (\ref{eq:s tilde}) and its corresponding equations in the other shift directions, we obtain (\ref{eq:ellKP2}). As referred to earlier, this representation of the system was chosen, despite containing the extra variables $U_{0,1}$ and $U_{1,0}$, as it clearly demonstrates the dependence on the elliptic moduli. If we now eliminate these extra variables, equations (\ref{eq:curve1},\ref{eq:curve2} and \ref{eq:curve3}) may be replaced in the system by the following three equations:
\begin{subequations}\label{eq:no e}
\begin{align}
&\left(p+\overline{u}-\frac{\widetilde{\overline{w}}}{\widetilde{\overline{s}}}\right)\left(p-\widetilde{u}-\frac{\overline{w}}{\overline{s}}\right)-
\left(q+\overline{u}-\frac{\widehat{\overline{w}}}{\widehat{\overline{s}}}\right)\left(q-\widehat{u}-\frac{\overline{w}}{\overline{s}}\right)\nonumber\\
&\hspace{30mm}=\frac{1}{\widehat{\overline{s}}s'}-\frac{1}{\widetilde{\overline{s}}s'}+g\widetilde{\widehat{s}}'\left(\widehat{s}-\widetilde{s}\right)+\left(p+q+\overline{u}-\widehat{\widetilde{u}}\right)\left(p-q+\widehat{u}-\widetilde{u}\right),\\
&\left(q+\overline{u}-\frac{\widehat{\overline{w}}}{\widehat{\overline{s}}}\right)\left(q-\widehat{u}-\frac{\overline{w}}{\overline{s}}\right)-
\left(r+\overline{u}-\frac{\dot{\overline{w}}}{\dot{\overline{s}}}\right)\left(r-\dot{u}-\frac{\overline{w}}{\overline{s}}\right)\nonumber\\
&\hspace{30mm}=\frac{1}{\dot{\overline{s}}s'}-\frac{1}{\widehat{\overline{s}}s'}+g\widehat{\dot{s}}'\left(\dot{s}-\widehat{s}\right)+\left(q+r+\overline{u}-\dot{\widehat{u}}\right)\left(q-r+\dot{u}-\widehat{u}\right),\\
&\left(r+\overline{u}-\frac{\dot{\overline{w}}}{\dot{\overline{s}}}\right)\left(r-\dot{u}-\frac{\overline{w}}{\overline{s}}\right)-
\left(p+\overline{u}-\frac{\widetilde{\overline{w}}}{\widetilde{\overline{s}}}\right)\left(q-\widetilde{u}-\frac{\overline{w}}{\overline{s}}\right)\nonumber\\
&\hspace{30mm}=\frac{1}{\widetilde{\overline{s}}s'}-\frac{1}{\dot{\overline{s}}s'}+g\dot{\widetilde{s}}'\left(\widetilde{s}-\dot{s}\right)+\left(r+p+\overline{u}-\widetilde{\widehat{u}}\right)\left(r-p+\widetilde{u}-\dot{u}\right).
\end{align}
\end{subequations}
This results in a closed-form elliptic lattice system that can be expressed only in terms of the variables $u$, $s$, $s'$ and $w$. A dual system, similar to (\ref{eq:Lattice KP}), in terms of the variables $h$, $v$, $s$ and $s'$ can also be derived, shadowing the system.

\section{Discrete Lax Representation}\label{discrete lax}
\setcounter{equation}{0}
To obtain a Lax Representation, we require the infinite component vectors, $\boldsymbol{u}_k$, introduced in (\ref{eq:u vector}), which can be rewritten as
\begin{equation}
\boldsymbol{u}_k\equiv\left(\boldsymbol{1}-\boldsymbol{U}\boldsymbol{\Omega}\right)\rho_k\boldsymbol{c}_k.
\end{equation} 
For these vectors the following set of shift relations can be derived:
\begin{subequations}
\begin{align}
&\widetilde{\boldsymbol{u}}_k=\left(p+\boldsymbol{\Lambda}\right)\boldsymbol{u}_k-\widetilde{\boldsymbol{U}}\left(\boldsymbol{O}-g{}^t\boldsymbol{L}^{-1}\boldsymbol{O}\boldsymbol{L}^{-1}\right)\boldsymbol{u}_k,\\
&\widehat{\boldsymbol{u}}_k=\left(q+\boldsymbol{\Lambda}\right)\boldsymbol{u}_k-\widehat{\boldsymbol{U}}\left(\boldsymbol{O}-g{}^t\boldsymbol{L}^{-1}\boldsymbol{O}\boldsymbol{L}^{-1}\right)\boldsymbol{u}_k,\\
&\dot{\boldsymbol{u}}_k=\left(r+\boldsymbol{\Lambda}\right)\boldsymbol{u}_k-\dot{\boldsymbol{U}}\left(\boldsymbol{O}-g{}^t\boldsymbol{L}^{-1}\boldsymbol{O}\boldsymbol{L}^{-1}\right)\boldsymbol{u}_k,\\
&\overline{\boldsymbol{u}}_k=-\boldsymbol{L}\boldsymbol{u}_k+\overline{\boldsymbol{U}}\left(\boldsymbol{O}\boldsymbol{\Lambda}-{}^t\boldsymbol{\Lambda}\boldsymbol{O}\right)\boldsymbol{u}_k,\\
&\boldsymbol{u}_k=-\boldsymbol{L}^{-1}\overline{\boldsymbol{u}}_k+\boldsymbol{U}{}^t\boldsymbol{L}^{-1}\left(\boldsymbol{O}\boldsymbol{\Lambda}-{}^t\boldsymbol{\Lambda}\boldsymbol{O}\right)\boldsymbol{L}^{-1}\overline{\boldsymbol{u}}_k.
\end{align}
\end{subequations}
Setting $\left(\boldsymbol{u}_k\right)_i=\varphi_i$, and introducing the 2-component vector $\boldsymbol{\varphi}=\left(\varphi_1, \varphi_2\right)^T$, we can derive the following Lax triplet:
\begin{subequations}\label{eq:lax triplet}
\begin{align}
&\widetilde{\boldsymbol{\varphi}}=A_0\boldsymbol{\varphi}+A_1\underline{\boldsymbol{\varphi}}+J\overline{\boldsymbol{\varphi}},\\
&\widehat{\boldsymbol{\varphi}}=B_0\boldsymbol{\varphi}+B_1\underline{\boldsymbol{\varphi}}+J\overline{\boldsymbol{\varphi}},\\
&\dot{\boldsymbol{\varphi}}=C_0\boldsymbol{\varphi}+C_1\underline{\boldsymbol{\varphi}}+J\overline{\boldsymbol{\varphi}},
\end{align}
\end{subequations}
where
\begin{subequations}
\begin{align}
&A_0=\left(\begin{array}{cc}
p-\widetilde{u} & 1\\
3e-\overline{U}_{0,1}-\widetilde{U}_{1,0} & p+\overline{u}\\
\end{array}\right),\\
&A_1=g\left(\begin{array}{cc}
-\widetilde{s}'w & \widetilde{s}'s\\
-\widetilde{w}'w & \widetilde{w}'s\\
\end{array}\right),\\
&J=\left(\begin{array}{cc}
0 & 0\\
-1 & 0\\
\end{array}\right),
\end{align}
\end{subequations}
with $B_i$ and $C_i$, $i=0,1$, equivalent to $A_i$, but with $p$ and $\widetilde{\cdot}$ shifts, replaced by $q$ and $r$ and $\widehat{\cdot}$ and $\dot{\cdot}$   shifts, respectively. The system (\ref{eq:lax triplet}) is subject to a number of pairwise compatibility relations, for example, between the $\widetilde{\cdot}$ and $\widehat{\cdot}$ directions resulting in the condition
\begin{equation}
\begin{split}
&\left(\widehat{A}_0J+J\overline{B}_0-\widetilde{B}_0J-J\overline{A}_0\right)\overline{\boldsymbol{\varphi}}
+\left(\widehat{A}_0B_0+\widehat{A}_1J+J\overline{B}_1-\widetilde{B}_0A_0-\widetilde{B}_1J-J\overline{A}_1\right)\boldsymbol{\varphi}\\
&\hspace{20mm}+\left(\widehat{A}_0B_1+\widehat{A}_1\underline{B}_0-\widetilde{B}_0A_1-\widetilde{B}_1\underline{A}_0\right)\underline{\boldsymbol{\varphi}}
+\left(\widehat{A}_1\underline{B}_1-\widetilde{B}_1\underline{A}_1\right)\underline{\underline{\boldsymbol{\varphi}}}=0.
\end{split}
\end{equation}
This compatibility condition results in the following system of eight equations:
\begin{subequations}
\begin{align}
&p\left(\widehat{\widetilde{u}}-\widehat{u}\right)-q\left(\widehat{\widetilde{u}}-\widetilde{u}\right)+\widetilde{U}_{1,0}-\widehat{U}_{1,0}+g\widehat{\widetilde{s}}'\left(\widetilde{s}-\widehat{s}\right)+\widehat{\widetilde{u}}\left(\widehat{u}-\widetilde{u}\right)=0,\\
&\left(q-\widehat{u}\right)\left(3e-\widehat{\overline{U}}_{0,1}-\widehat{\widetilde{U}}_{1,0}\right)+\left(p+\widehat{\overline{u}}\right)\left(3e-\overline{U}_{0,1}-\widehat{U}_{1,0}\right)\nonumber\\
&\hspace{20mm}-\left(p-\widetilde{u}\right)\left(3e-\widetilde{\overline{U}}_{0,1}-\widetilde{\widehat{U}}_{1,0}\right)-\left(q+\widetilde{\overline{u}}\right)\left(3e-\overline{U}_{0,1}-\widetilde{U}_{1,0}\right)\nonumber\\
&\hspace{80mm}=g\left(\widehat{\widetilde{w}}'\left(\widehat{s}-\widetilde{s}\right)+\overline{w}\left(\widetilde{\overline{s}}'-\widehat{\overline{s}}'\right)\right),\\
&p\left(u-\widetilde{u}\right)-q\left(u-\widehat{u}\right)+\widetilde{U}_{0,1}-\widehat{U_{0,1}}+gs\left(\widetilde{s}'-\widehat{s}'\right)+u\left(\widehat{u}-\widetilde{u}\right)=0,\\
&g\left(\widehat{s}'w\left(p-\widehat{\widetilde{u}}\right)+\widehat{w}'w+\widehat{\widetilde{s}}'\widehat{w}\left(q-\underline{\widehat{u}}\right)-\widehat{\widetilde{s}}'\widehat{s}\left(3e-U_{0,1}-\underline{\widehat{U}}_{1,0}\right)\right.\nonumber\\
&\hspace{20mm}\left.-\widetilde{s}'w\left(q-\widehat{\widetilde{u}}\right)-\widetilde{w}'w-\widehat{\widetilde{s}}'\widehat{w}\left(p-\underline{\widetilde{u}}\right)+\widehat{\widetilde{s}}'\widehat{s}\left(3e-U_{0,1}-\underline{\widetilde{U}}_{1,0}\right)\right)=0,\\
&g\left(\widehat{s}'s\left(p-\widetilde{\widehat{u}}\right)+\widehat{w}'s-\widehat{\widetilde{s}}'\widehat{w}+\widehat{\widetilde{s}}'\widehat{s}\left(q+u\right)\right.\nonumber\\
&\left.\hspace{60mm}-\widetilde{s}'s\left(q-\widetilde{\widehat{u}}\right)-\widetilde{w}'s+\widehat{\widetilde{s}}'\widetilde{w}-\widehat{\widetilde{s}}'\widetilde{s}\left(p+u\right)\right)=0,\\
&g\left(-\widehat{s}'w\left(3e-\widehat{\overline{U}}_{0,1}-\widehat{\widetilde{U}}_{0,1}\right)-\widehat{w}'w\left((p+\widehat{\overline{u}}\right)
+\widehat{s}\widehat{\widetilde{w}}'\left(3e-U_{0,1}-\widehat{\underline{U}}_{0,1}\right)\right.\nonumber\\
&\left.\hspace{20mm}-\widetilde{\widehat{w}}'\widehat{w}\left((q-\widehat{\underline{u}}\right)
+\widetilde{s}'w\left(3e-\widetilde{\overline{U}}_{0,1}-\widehat{\widetilde{U}}_{0,1}\right)+\widetilde{w}'w\left((q+\widetilde{\overline{u}}\right)\right.\nonumber\\
&\left.\hspace{40mm}-\widetilde{s}\widehat{\widetilde{w}}'\left(3e-U_{0,1}-\widetilde{\underline{U}}_{0,1}\right)+\widetilde{\widehat{w}}'\widetilde{w}\left((p-\widetilde{\underline{u}}\right)\right)=0,\\
&g\left(\widehat{s}'s\left(3e-\widehat{\overline{U}}_{0,1}-\widehat{\widetilde{U}}_{0,1}\right)+\widehat{w}'s\left(p+\widehat{\overline{u}}\right)
-\widehat{\widetilde{w}}'\widehat{w}+\widehat{\widetilde{w}}'\widehat{s}\left(q+u\right)\right.\nonumber\\
&\left.\hspace{20mm}-\widetilde{s}'s\left(3e-\widetilde{\overline{U}}_{0,1}-\widetilde{\widehat{U}}_{0,1}\right)-
\widetilde{w}'s\left(q+\widetilde{\overline{u}}\right)
+\widehat{\widetilde{w}}'\widetilde{w}-\widehat{\widetilde{w}}'\widetilde{s}\left(p+u\right)\right)=0,\\
&s'\overline{w}=w'\overline{s}.
\end{align}
\end{subequations}
Similar compatibility conditions also exist between the $\widetilde{\cdot}$ and $\dot{\cdot}$ shifts, and the $\widehat{\cdot}$ and $\dot{\cdot}$ shifts. The system (\ref{eq:Lattice KP}) then follows from these compatibility conditions.

\section{Dimensional Reduction and Degeneration}

\setcounter{equation}{0}
Both the lattice, and continuous, Potential KP equation and the Elliptic KdV equations may be recovered under particular limits of the systems (\ref{eq:Lattice KP}), and (\ref{eq:continuousKP}), respectively.
\subsection{Degeneration of the Elliptic Curve}

Taking the limit $g\to0$ causes the elliptic curve to degenerate, and from (\ref{eq:LPKP}) we recover the lattice potential KP equation\cite{LPKP},
\begin{equation}
\left(p-\widetilde{u}\right)\left(q-r+\widetilde{\dot{u}}-\widetilde{\widehat{u}}\right)+
\left(q-\widehat{u}\right)\left(r-p+\widehat{\widetilde{u}}-\widehat{\dot{u}}\right)+
\left(r-\dot{u}\right)\left(p-q+\dot{\widehat{u}}-\dot{\widetilde{u}}\right)
=0,
\end{equation}
which first appeared in \cite{LPKP2}. In contrast to the bilinear lattice KP of \cite{DAGTE}, this has a continuum limit to the actual potential KP equation.

In the continuous case, taking the limit $g=0$ in equation (\ref{eq:KP1}), we recover the potential KP equation,
\begin{equation}
\left(u_t-\frac{1}{4}u_{xxx}-\frac{3}{2}\left(u_x\right)^2\right)_x=\frac{3}{4}u_{yy}.
\end{equation}

\subsection{Dimensional Reduction}
Both the elliptic lattice and continuous KdV systems, given in \cite{ellipticKdv} require that the matrices $\boldsymbol{C}$, and hence $\boldsymbol{U}$ are symmetric under transposition,
\begin{equation}
\,\phantom{a}^{t\!}\boldsymbol{C}=\boldsymbol{C} \Rightarrow \,\,\phantom{a}^{t\!}\boldsymbol{U}=\boldsymbol{U},
\end{equation}
i.e. $U_{i,j}=U_{j,i}$. This restriction ensures that the primed, and bar shifted, variables become equal to their unprimed, and unbarred, equivalents, respectively,
\begin{equation}
\begin{array}{ccc}
s'\to s, & v'\to v & w'\to w,\\
\overline{u}\to u, & \overline{s}\to s & \overline{w}\to w,\\
\end{array}
\end{equation}
and results in  a dimensional reduction, from which the elliptic lattice KdV system is recovered from (\ref{eq:Lattice KP}), and the elliptic continuous KdV system is recovered from (\ref{eq:continuousKP}). 

\section{Soliton Type Solutions}\label{soliton solutions}
\setcounter{equation}{0}
As a concrete application of the infinite matrix scheme used to derive the lattice system, it is relatively straightforward to construct soliton type solutions. Introducing the $\mathcal{N}$ by $\mathcal{N}'$ matrix $M$, defined by
\begin{equation}
\boldsymbol{M}=\Omega(k_i,k'_j)\boldsymbol{r}{}^{t}\boldsymbol{s},
\end{equation}
with entries
\begin{equation}
M_{ij}=\frac{1-g/K_iK'_j}{k_i+k'_j}\rho_i\sigma_j,\hspace{10mm}(i=1,\dots,\mathcal{N};j=1,\dots,\mathcal{N}'),
\end{equation}
where the parameters of the solutions $(k_i,K_i)$ and $(k'_j,K'_j)$ are points on the elliptic curve $\Gamma$, (\ref{eq:elliptic curve}), and the vectors $\boldsymbol{r}$, $\,\phantom{a}^{t\!}\boldsymbol{s}$ are given by
\begin{subequations}
\begin{align}
\boldsymbol{r}&=\left(\begin{array}{c}
\rho_1\\
\vdots\\
\rho_{\mathcal{N}}
\end{array}\right),\\
\,\,\phantom{a}^{t\!}\boldsymbol{s}&=\left(\begin{array}{ccc}
\sigma_1, & \cdots & \sigma_{\mathcal{N}'}\\
\end{array}\right).
\end{align}
\end{subequations}
In order to obtain the soliton type solutions we take the infinite matrix $\boldsymbol{C}$ to be a finite rank, $\mathcal{N}'$ by $\mathcal{N}$ matrix. We also define diagonal matrices
\begin{subequations}
\begin{align}
&\boldsymbol{k}=\textrm{diag}(k_1,k_2,\dots,k_{\mathcal{N}}),\hspace{10mm}\boldsymbol{K}=\textrm{diag}(K_1,K_2,\dots,K_{\mathcal{N}}),\\
&\boldsymbol{k'}=\textrm{diag}(k'_1,k'_2,\dots,k'_{\mathcal{N}'}),\hspace{10mm}\boldsymbol{K'}=\textrm{diag}(K'_1,K'_2,\dots,K'_{\mathcal{N}'}).
\end{align}
\end{subequations}
This leads to the following explicit solutions:
\begin{subequations}
\begin{align}
&u=\,\,\phantom{a}^{t\!}\boldsymbol{s}(\boldsymbol{1}+\boldsymbol{C}\boldsymbol{M})^{-1}\boldsymbol{C}\boldsymbol{r},\\
&s=\,\,\phantom{a}^{t\!}\boldsymbol{s}\boldsymbol{K}'^{-1}(\boldsymbol{1}+\boldsymbol{C}\boldsymbol{M})^{-1}\boldsymbol{C}\boldsymbol{r},\\
&s'=\,\,\phantom{a}^{t\!}\boldsymbol{s}(\boldsymbol{1}+\boldsymbol{C}\boldsymbol{M})^{-1}\boldsymbol{C}\boldsymbol{K}^{-1}\boldsymbol{r},\\
&h=\,\,\phantom{a}^{t\!}\boldsymbol{s}\boldsymbol{K}'^{-1}(\boldsymbol{1}+\boldsymbol{C}\boldsymbol{M})^{-1}\boldsymbol{C}\boldsymbol{K}^{-1}\boldsymbol{r},\\
&v=1-\,\phantom{a}^{t\!}\boldsymbol{s}\boldsymbol{K}'^{-1}\boldsymbol{k}'(\boldsymbol{1}+\boldsymbol{C}\boldsymbol{M})^{-1}\boldsymbol{C}\boldsymbol{r},\\
&v'=1-\,\phantom{a}^{t\!}\boldsymbol{s}(\boldsymbol{1}+\boldsymbol{C}\boldsymbol{M})^{-1}\boldsymbol{C}\boldsymbol{k}\boldsymbol{K}^{-1}\boldsymbol{r},\\
&w=1+\,\phantom{a}^{t\!}\boldsymbol{s}\boldsymbol{K}'^{-1}(\boldsymbol{1}+\boldsymbol{C}\boldsymbol{M})^{-1}\boldsymbol{C}\boldsymbol{k}\boldsymbol{r},\\
&w'=1+\,\phantom{a}^{t\!}\boldsymbol{s}\boldsymbol{k}'(\boldsymbol{1}+\boldsymbol{C}\boldsymbol{M})^{-1}\boldsymbol{C}\boldsymbol{K}^{-1}\boldsymbol{r}.
\end{align}
\end{subequations}
Note, that although the dynamics do not explicitly involve the elliptic curve, the soliton solutions are essentially dependent on the variables on the curve. 

\section{Associated Continuous Systems}
\setcounter{equation}{0}
Here we concentrate on the derivation of the analogous continuous system to the elliptic discrete lattice KP system (\ref{eq:LPKP}). Many discrete integrable lattice systems possess such compatible continuous systems, with these forming continuous symmetries for the lattice systems, whilst in turn these lattice systems constitute discrete symmetries for the corresponding continuous flows. We provide some of the simplest of these continuous flows for purpose of identification with the associated lattice system. 

For the continuous case the plane wave factors are taken to be continuous exponential functions, given by
\begin{subequations}
\begin{align}
\rho(k)=\textrm{exp}\left(\sum_{j\in\mathbb{Z}}k^jx_j\right),\\
\sigma(k')=\textrm{exp}\left(-\sum_{j\in\mathbb{Z}}\left(-k'\right)^jx_j\right),
\end{align}
\end{subequations}
yielding
\begin{equation}\label{eq:deriv}
\frac{\partial\boldsymbol{U}}{\partial x_i}=\boldsymbol{\Lambda}^i\boldsymbol{U}-\boldsymbol{U}\left(-\,\,\phantom{a}^{t\!}\boldsymbol{\Lambda}\right)^i+\boldsymbol{U}\boldsymbol{O}_i\boldsymbol{U},
\end{equation}
where $\boldsymbol{O}_i$ was given in (\ref{eq:Ojhat}). From (\ref{eq:deriv}), we can then derive
and hence,
\begin{equation}\label{eq:U i+j deriv}
\frac{\partial}{\partial x_{i+j}}\boldsymbol{U}=\left(\boldsymbol{\Lambda}^j-\boldsymbol{U}\widehat{\boldsymbol{O}}_j\right)\left(\frac{\partial}{\partial x_{i}}\boldsymbol{U}\right)+\left(\frac{\partial}{\partial x_{j}}\boldsymbol{U}\right)\left(\left(-\,\,\phantom{a}^{t\!}\boldsymbol{\Lambda}\right)^i+\widehat{\boldsymbol{O}}_i\boldsymbol{U}\right).
\end{equation}
Denoting $x=x_1$, $y=x_2$ and $t=x_3$, it follows that the basic relations for the corresponding flows are
\begin{subequations}
\begin{align}
&\boldsymbol{U}_x=\boldsymbol{\Lambda}\boldsymbol{U}+\boldsymbol{U}\,\,\phantom{a}^{t\!}\boldsymbol{\Lambda}-\boldsymbol{U}\widehat{\boldsymbol{O}}_1\boldsymbol{U},\label{eq:x-deriv}\\
&\boldsymbol{U}_y=\left(\boldsymbol{\Lambda}\boldsymbol{U}-\boldsymbol{U}\,\,\phantom{a}^{t\!}\boldsymbol{\Lambda}\right)_x+\boldsymbol{U}_x\widehat{\boldsymbol{O}}_1\boldsymbol{U}-\boldsymbol{U}\widehat{\boldsymbol{O}}_1\boldsymbol{U}_x,\\
&\left(\boldsymbol{U}_t-\frac{1}{4}\boldsymbol{U}_{xxx}-\frac{3}{2}\boldsymbol{U}_x\widehat{\boldsymbol{O}}_1\boldsymbol{U}_x\right)_x=\frac{3}{4}\boldsymbol{U}_{yy}+\frac{3}{2}\left(\boldsymbol{U}_y\widehat{\boldsymbol{O}}_1\boldsymbol{U}_x-\boldsymbol{U}_x\widehat{\boldsymbol{O}}_1\boldsymbol{U}_y\right),
\end{align}
\end{subequations}
together with (\ref{eq:barshift1}) and (\ref{eq:barshift2}), from which we derive for the $x$-derivatives
\begin{subequations}
\begin{align}
&u_x=U_{1,0}+U_{0,1}-u^2+gss',\\
&-\left(u-gh\right)=\frac{s_x+v-w}{s}=\frac{s'_x+v'-w'}{s'},\\
&h_x=U_{-1,-2}+U_{-2,-1}-ss'+gh^2,\\
&-v_x=v\left(u+g\underline{h}\right)+U_{-1,1}+gs\left(U_{-1,-2}+\underline{U}_{-2,-1}\right)+3es,\\
&-v'_x=v'\left(u+g\overline{h}\right)+U_{1,-1}+gs'\left(\overline{U}_{-1,-2}+U_{-2,-1}\right)+3es',\\
&w_x=w\left(\underline{u}+gh\right)+U_{-1,1}-s\left(U_{0,1}+\underline{U}_{1,0}\right)+3es,\\
&w'_x=w'\left(\overline{u}+gh\right)+U_{1,-1}-s'\left(\overline{U}_{0,1}+U_{1,0}\right)+3es'.
\end{align}
\end{subequations}
As for the $y$-derivatives, we have
\begin{subequations}
\begin{align}
&u_y=\left(U_{1,0}-U_{0,1}\right)_x+g\left(s's_x-ss'_x\right),\\
&h_y=\left(U_{-1,-2}-U_{-2,-1}\right)_x+g\left(ss'_x-s's_x\right),\\
&s_y=-\left(v+w\right)_x+s_x\left(u+gh\right)-s\left(u+gh\right)_x,\\
&s'_y=\left(v'+w'\right)_x+s'\left(u+gh\right)_x-s'_x\left(u+gh\right),\\
&v_y=\left(U_{-1,1}\right)_x-\left(vu_x+uv_x\right)+g\left(s\left(U_{-1,-2}\right)_x-U_{-1,-2}s_x\right),\\
&v'_y=-\left(U_{1,-1}\right)_x+\left(uv'_x+v'u_x\right)+g\left(\left(U_{-2,-1}\right)s'_x-s'\left(U_{-2,-1}\right)_x\right),\\
&w_y=\left(U_{-1,1}-U_{-2,2}\right)_x+\left(U_{0,1}s_x-s\left(U_{0,1}\right)_x\right)+g\left(hw_x-h_x\left(w-1\right)\right),\\
&w'_y=\left(U_{2,-2}-U_{1,-1}\right)_x+\left(U_{0,1}s'_x-s'\left(U_{0,1}\right)_x\right)+g\left(\left(w'-1\right)h_x-hw'_x\right),\\
&u_y+u_{xx}=2\left(U_{1,0}\right)_x-2uu_x+2gs'h_x,\label{eq u10 diff}\\
&u_y-u_{xx}=-2\left(U_{0,1}\right)_x+2uu_x-2gsh_x.\label{eq u01 diff}
\end{align}
\end{subequations}
Whilst, for the $t$-derivatives, 
\begin{subequations}
\begin{align}
&\left(u_t-\frac{1}{4}u_{xxx}-\frac{3}{2}\left(u_x\right)^2+\frac{3}{2}gs_xs'_x\right)_x=\frac{3}{4}u_{yy}+\frac{3}{2}g\left(s'_xs_y-s'_ys_x\right),\label{eq:u_t}\\
&\left(h_t-\frac{1}{4}h_{xxx}-\frac{3}{2}s_xs'_x+\frac{3}{2}g\left(h_x\right)^2\right)_x=\frac{3}{4}h_{yy}+\frac{3}{2}\left(s'_xs_y-s'_ys_x\right),\label{eq:h_t}\\
&\left(s_t-\frac{1}{4}s_{xxx}-\frac{3}{2}s_xu_x+\frac{3}{2}gh_xs_x\right)_x=\frac{3}{4}s_{yy}+\frac{3}{2}\left(s_y\left(u+gh\right)_x-s_x\left(u+gh\right)_y\right),\label{eq:s_t}\\
&\left(s'_t-\frac{1}{4}s'_{xxx}-\frac{3}{2}s'_xu_x+\frac{3}{2}gh_xs'_x\right)_x=\frac{3}{4}s'_{yy}+\frac{3}{2}\left(s'_x\left(u+gh\right)_y-s'_y\left(u+gh\right)_x\right)\label{eq:s'_t},\\
&\left(v_t-\frac{1}{4}v_{xxx}-\frac{3}{2}v_xu_x+\frac{3}{2}g\left(U_{-1,-2}\right)_xs_x\right)_x\nonumber\\
&\hspace{20mm}=\frac{3}{4}v_{yy}+\frac{3}{2}\left(v_yu_x-v_xu_y\right)+\frac{3}{2}g\left(\left(U_{-1,-2}\right)_xs_y-\left(U_{-1,-2}\right)_ys_x\right),\\
&\left(v'_t-\frac{1}{4}v'_{xxx}-\frac{3}{2}v'_xu_x-\frac{3}{2}g\left(U_{-2,-1}\right)_xs'_x\right)_x\nonumber\\
&\hspace{20mm}=\frac{3}{4}v'_{yy}+\frac{3}{2}\left(u_yv'_x-u_xv'_y\right)+\frac{3}{2}g\left(\left(U_{-2,-1}\right)_ys_x-\left(U_{-2,-1}\right)_xs_y\right),\\
&\left(w_t-\frac{1}{4}w_{xxx}-\frac{3}{2}s_x\left(U_{0,1}\right)_x+\frac{3}{2}gh_xw_x\right)_x\nonumber\\
&\hspace{20mm}=\frac{3}{4}w_{yy}++\frac{3}{2}\left(s_y\left(U_{0,1}\right)_x-s_x\left(U_{0,1}\right)_y\right)+\frac{3}{2}g\left(h_xw_y-h_yw_x\right),\\
&\left(w_t'-\frac{1}{4}w'_{xxx}-\frac{3}{2}s'_x\left(U_{1,0}\right)_x+\frac{3}{2}gh_xw'_x\right)_x\nonumber\\
&\hspace{20mm}=\frac{3}{4}w'_{yy}++\frac{3}{2}\left(s'_x\left(U_{1,0}\right)_y-s_y\left(U_{1,0}\right)_x\right)+\frac{3}{2}g\left(h_yw'_x-h_xw'_y\right).
\end{align}
\end{subequations}
The closed system, (\ref{eq:continuousKP}),in terms of the $u$, $s$, $s'$ and $w$  results by eliminating all other variables, except $U_{0,1}$ and $U_{1,0}$ which can be eliminated by differentiating both equations (\ref{eq:curve dep1}) and (\ref{eq:curve dep2}) with respect to $x$, and substituting in equations (\ref{eq u10 diff}) and (\ref{eq u01 diff}), respectively.

\section{Continuous Lax Representation}
\setcounter{equation}{0}
This system is integrable by construction, and in fact admits soliton solutions of the same form as in section \ref{soliton solutions}, but with $\rho$ and $\sigma$replaced by their continuous analogues. The integrability can be made more apparent by the fact that it admits a Lax triplet, given by
\begin{subequations}
\begin{align}
\boldsymbol{\varphi}_y&=\boldsymbol{\varphi}_{xx}+A\boldsymbol{\varphi}+B\underline{\boldsymbol{\varphi}},\\
\boldsymbol{\varphi}_x&=J\overline{\boldsymbol{\varphi}}+C\boldsymbol{\varphi}+D\underline{\boldsymbol{\varphi}},\label{x lax}\\
\boldsymbol{\varphi}_t&=\boldsymbol{\varphi}_{xy}+E\boldsymbol{\varphi}+F\boldsymbol{\varphi}_x+G\underline{\boldsymbol{\varphi}},
\end{align}
\end{subequations}
with $\boldsymbol{\varphi}$ as in section \ref{discrete lax}, and where
\begin{subequations}
\begin{align}
&J=\left(\begin{array}{cc}
0 & 0\\
1 & 0
\end{array}\right)\\
&A=\left(\begin{array}{cc}
2u_x & 0\\
2\left(U_{1,0}\right)_x & 0\\
\end{array}\right),\\
&B=\left(\begin{array}{cc}
-2gs'_xw & 2gs'_xs\\
-2gw'_xw & 2gw'_xs\\
\end{array}\right),\\
&C=\left(\begin{array}{cc}
-u & 1\\
3e-U_{1,0}-\overline{U}_{0,1} & \overline{u}\\
\end{array}\right),\\
&D=\left(\begin{array}{cc}
gws' & -gss'\\
gww' & -gsw'\\
\end{array}\right),\\
&E=\left(\begin{array}{cc}
u_y-\left(U_{0,1}\right)_x-\frac{u\left(s'_y+v'_x\right)}{s'} & u_x+\frac{s'_y+v'_x}{s'}\\
\left(U_{1,0}\right)_y-\left(U_{1,1}\right)_x-\frac{u\left(\left(U_{1,-1}\right)_x-w'_y\right)}{s'} & \left(U_{1,0}\right)_x+\frac{v'_y-\left(U_{1,-1}\right)_x}{s'}
\end{array}\right),\\
&F=\left(\begin{array}{cc}
\frac{-s'_y-v'_x}{s'} & 0\\
\frac{\left(U_{1,-1}\right)_x-w'_y}{s'} & 0\\
\end{array}\right),\\
&G=\left(\begin{array}{cc}
-gs'_xU_{1,-1} & -gs'_xv\\
-gw'_xU_{-1,1} & -gw'_xv
\end{array}\right).
\end{align}
\end{subequations}
Again, all of the other variables, except $u$, $s$, $s'$ and $w$ and their derivatives must be eliminated, and the system (\ref{eq:continuousKP}) follows from the compatibility between $\boldsymbol{\varphi}_y$ and $\boldsymbol{\varphi}_t$, with all $x$-derivatives replaced by $\overline{\cdot}$ shifts using (\ref{x lax}).

\section{Conclusion}
We have derived from a direct linearisation scheme a $3+1$ dimensional lattice system, naturally associated with an elliptic curve, as an extension of the lattice potential KP equation. We have also shown this system to be integrable through the existence of a Lax representation and soliton solutions. To our knowledge, this is the first lattice system associated with an elliptic curve in higher dimensions proposed. An analogous continuous system with three continuous variables and one discrete variable was also derived. This continuous system is reminiscent to, but different from, an elliptic generalisation of the KP equation given in \cite{Date/Jimbo/Miwa}. Our understanding is that the difference is in the choice of Cauchy kernel. Whereas our system has a Cauchy kernel related to the elliptic KdV system, we believe that \cite{Date/Jimbo/Miwa} has a Cauchy kernel related to the Landau-Lifshitz equation. We intend to investigate the correspondence between the two cases in future.

\section*{Acknowledgments}
Paul Jennings is supported by the UK Engineering and Physical Sciences Research Council (EPSRC).

\section*{References}
\bibliography{Elliptic_KP}
\bibliographystyle{unsrt}

\end{document}